\documentclass[doublecol]{epl2} 

\usepackage[breaklinks=true,colorlinks=true,linkcolor=blue,urlcolor=blue,citecolor=blue]{hyperref}

\usepackage[mathscr]{euscript}
\usepackage[ruled,linesnumbered]{algorithm2e}
\usepackage{amsmath,amssymb,amscd}
\usepackage[T1]{fontenc}
\usepackage{booktabs}
\usepackage{epigraph}

\title{How Committed Individuals Shape Social Dynamics: A Survey on Coordination Games and Social Dilemma Games}
\shorttitle{A survey on the committed individuals} 

\author{Chen Shen\inst{1} \and Hao Guo\inst{2,3} \and Shuyue Hu\inst{4} \and Lei Shi\inst{5} \thanks{Email: \email{shi\_lei65@hotmail.com}} \and Zhen Wang\inst{3} \thanks{Email: \email{w-zhen@nwpu.edu.cn}} 
\and Jun Tanimoto\inst{1,6} }
\shortauthor{C. Shen \etal}

\institute{                    
  \inst{1} Faculty of Engineering Sciences, Kyushu University, Fukuoka, 816-8580, Japan\\
  \inst{2} School of Mechanical Engineering, Northwestern Polytechnical University, Xi'an {\rm 710072}, China\\
   \inst{3} School of Artificial Intelligence, Optics and Electronics (iOPEN), Northwestern Polytechnical University, Xi'an {\rm 710072}, China\\
      \inst{4} Shanghai Artificial Intelligence Laboratory, Shanghai, China\\
    \inst{5} School of Statistics and Mathematics, Yunnan University of Finance and Economics, 650221, Kunming, China\\
     \inst{6} Interdisciplinary Graduate School of Engineering Sciences, Kyushu University, Fukuoka, 816-8580, Japan
}
\abstract{
Committed individuals, who features steadfast dedication to advocating strong beliefs, values, and preferences, have garnered much attention across statistical physics, social science, and computer science. This survey delves into the profound impact of committed individuals on social dynamics that emerge from coordination games and social dilemma games. Through separate examinations of their influence on coordination, including social conventions and color coordination games, and social dilemma games, including one-shot settings, repeated settings, and vaccination games, this survey reveals the significant role committed individuals play in shaping social dynamics. Their contributions range from accelerating or overturning social conventions to addressing cooperation dilemmas and expediting solutions for color coordination and vaccination issues. Furthermore, the survey outlines three promising directions for future research: conducting human behavior experiments for empirical validation, leveraging advanced large language models as proxies for committed individuals in complex scenarios, and addressing potential negative impacts of committed individuals.}

\begin{document}

\maketitle

\epigraph{\itshape Never doubt that a small group of thoughtful, committed people can change the world; indeed, it's the only thing that ever has.}
{---Margaret Mead}

\vspace{-0.15in}
\section{Introduction}
Committed individuals (or people) are those who firmly hold their beliefs, values, and preferences, demonstrating unwavering dedication to promoting their convictions. These individuals, found in various domains, both in the physical and digital worlds, can have a profound impact on social dynamics.
Greta Thunberg, who is known for her faithful activism on climate change, has increased public awareness of environmental issues.
Tarana Burke, who started and tirelessly campaigned for the \#MeToo movement, has sparked a global conversation on sexual harassment and assault, leading to significant societal changes.
In the digital realm, social media bots, which are  programmed to constantly express certain opinions, are found to influence or even distort public opinions in the Catalan
independence referendum \cite{stella2018bots}, the Brexit referendum \cite{bessi2016social}, and the US presidential election \cite{vosoughi2018spread}. 

How do committed individuals shape social dynamics?
Under what conditions will they result in significant societal changes? These questions have piqued much interest across multiple disciplines, ranging from statistical physics \cite{Galam2012SociophysicsAP,Mobilia2003does,redner2019reality} to social science \cite{cherry2019stubborn,centola2018experimental} and computer science \cite{morris2019norm,kearns2006experimental}.
Specifically within the study of opinion dynamics, the importance of committed individuals has long been acknowledged \cite{noorazar2020recent,galam2012sociophysics,redner2019reality}, with research on this topic dating back to 1990s \cite{galam1991towards}.
More recently, in parallel to opinion dynamics, an emerging line of research examines the role that committed individuals play in social dynamics emerging from coordination games and social dilemma games \cite{centola2018experimental,masuda2012evolution}.
Adopting a game-theoretic framework to model microscopic interactions between individuals, this line of research assumes that individuals will receive payoffs during interactions and that the payoffs will depend on their jointly strategies.
As a result, typically, individuals are incentivized to develop their own strategies towards the direction of increasing their payoffs in interactions, rather than fusing the others' strategies to form their own as in the study of opinion dynamics~\cite{dong2018survey}.

This survey delves into the above two questions, with a specific focus on social dynamics that emerge from coordination games and social dilemma games. 
Coordination games involve individuals aligning their strategies to achieve a mutually preferred outcome. Social dilemma games model scenarios where individual benefits may conflict with collective benefits, and cooperation is necessary for collective well-being.
Existing studies have shown that committed individuals play a crucial role in various social dynamics, such as accelerating the emergence of social conventions from pure coordination games\cite{sen2007emergence,mukherjee2008norm,hao2013achieving}, overturning established and socially accepted social conventions, expediting the resolution of color coordination games~\cite{shirado2017locally}, 
triggering critical mass effects of cooperation emergence in social dilemmas ~\cite{cardillo2020critical,he2023impact},
promoting and sustaining cooperation
~\cite{PhysRevE.86.011134,masuda2012evolution}, 
and promoting vaccination behavior under social dilemmas coupled with epidemic processes ~\cite{liu2012impact,fukuda2016effects}.
Albeit significant progress along this line of research, so far there lacks a comprehensive survey which summarizes and discusses the progress centering around committed individuals.

The remainder of this survey will be organized as follows: First, we will summarize the impact of committed individuals on coordination games, focusing on their influence on social conventions and color coordination issues. Following that, we will shift our focus to social dilemma games, summarizing the influence of committed individuals in one-shot and anonymous games, exploring their impact in repeated social dilemma games, and reviewing their significance in vaccination dilemma games. Lastly, we will conclude with a summary of this survey and provide perspectives for future research in three key areas: conducting human behavior experiments to validate theoretical findings, leveraging advanced large language models as proxies for committed individuals in complex scenarios, and addressing potential negative impacts of committed individuals.

It is worth mentioning that the terminology for committed individuals varies across different fields. In opinion dynamics research, they are termed `inflexible agents' or `stubborn agents', while in coordination games, they are `non-learning agents' or `fixed-strategy agents'. In social dilemma games, they are referred to as `zealots' or `stubborn players'. However, these individuals all share a common principle: they consistently adhere to their beliefs or strategies and remain immune to external influence. For clarity, we will use the term `committed individuals' throughout this survey.

\section{Coordination game}
The coordination problem centers around aligning individual actions to achieve a mutually desired outcome. Typical examples of coordination problems include the pure coordination game where players benefit from aligning their choices, and network color game in which players try to avoid choosing the same option as others to maximize their individual gains.

\subsection{Pure coordination game}
The concept of committed individuals has gained significant attention in the field of coordination games, specifically regarding the emergence of social conventions~\cite{morris2019norm,haynes2017engineering}. Social conventions are widely accepted norms, rules, or behaviors that govern social interactions within a particular group, society, or culture. Researchers in this field aim to understand how social conventions are established, maintained, modified, and how they shape individual behavior and societal functioning\footnote{Although the terms `social convention' and `social norm' are sometimes used interchangeably, we will use the term `social convention' in accordance with the literature~\cite{haynes2017engineering}.}. Several theoretical studies have explored the potential impact of committed individuals on the emergence/convergence of social conventions by integrating them into different learning rules. For instance, Sen \etal~ proposed a social learning model where agents acquire their policies through repeated interactions with multiple agents, diverging from the traditional setting of learning from repeated interactions with the same player. By introducing a small number of committed agents who adopted a fixed-strategy, they demonstrated the ability of these agents to establish their actions as a global convention, influencing the behavior of other agents \cite{sen2007emergence}. Building upon Sen \etal's social learning rule, Mukherjee \etal~ investigated the influence of committed agents on the emergence of global social conventions on a grid. They found that when these agents adopt the same fixed strategy, the emergence of a social convention is expedited.  Conversely, when they employ different strategies, multiple social conventions can arise with equal frequency across multiple simulation runs. However, it's important to note that these multiple social conventions remain unobservable within the population in any single specific simulation.~\cite{mukherjee2008norm}. Within the reinforcement learning framework, Hao \etal~ examined the effects of committed agents on social optimal outcomes in asymmetric anti-coordination games, revealing that even a small number of committed agents can exert significant influence on the socially optimal outcome towards which the agents converge \cite{hao2013achieving}. Griffiths \etal~ examined how the introduction of committed individuals into various network topologies using the Q-learning algorithm affects the speed at which a global convention emerges\cite{griffiths2012impact}. Yu \etal~ assessed the impact of committed agents on the emergence of social conventions in networked multi-agent systems by employing a collective learning rule that combines several different learning rules. They found that emergence of a convention is unlikely in a society consisting solely of committed agents but is possible when there are also a small number of learners employing a fixed observation strategy \cite{yu2014collective}.  Additionally, Borglund \etal~ investigated the effects of committed agents on the emergence of local conventions and concluded that a larger number of committed agents generally leads to a faster emergence of local conventions, with only a few committed agents required to persuade communities to adopt fixed actions~\cite{borglund2018effects}. In their work, local conventions are specific agreements or patterns of behavior that emerge within smaller subgroups or communities, where nodes within the communities by definition are internally densely connected but externally sparsely connected. A mathematical definition of local conventions is found in ref.~\cite{hu2017achieving}.

\begin{figure}[!t]
\centering
\includegraphics[scale=0.61]{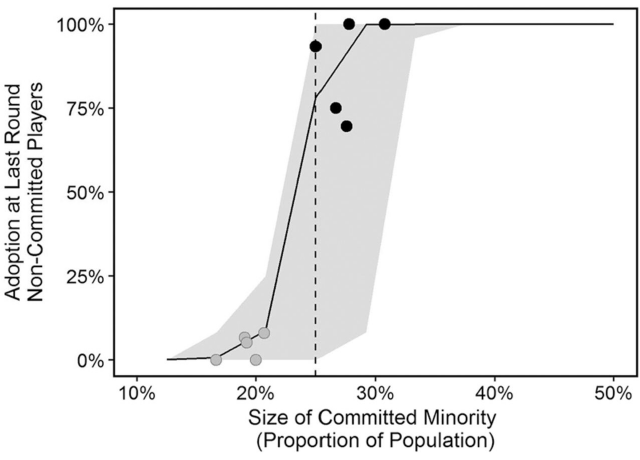}
\caption{Success Levels in overturning established social conventions across all trials. Gray circles represent trials with a critical size $C < 25\%$, while black circles represent trials with a critical size $C \geq 25\%$. The solid line corresponds to the theoretically predicted critical mass point at which committed individuals can successfully overturn the established social convention, with the gray area representing the 95\% confidence intervals. The dotted line represents the critical size $C = 25\%$. This figure is adapted from reference~\cite{centola2018experimental}.
}
\label{f02}
\end{figure}

In contrast to the aforementioned studies that primarily focused on how and why committed agents promote or accelerate the emergence of social conventions, there are also studies that explore other aspects. For instance, Marchant \etal~ demonstrated that committed agents can destabilize or eliminate established global conventions. They investigated the minimum intervention required for destabilization and analyzed the impact of pricing mechanisms on intervention costs \cite{marchant2014destabilising}. In addition, one recent study investigated how the committed minorities can overturn the established social convention through a large-scale human behavior experiment. In the study, Centola \etal~ adopted an experimental approach to studying tipping point dynamics within an artificially created system of evolving social conventions. 194 participants, who took part in a social coordination process. They were randomly assigned to one of 10 groups of 20-30 people. Participants were engaged in pairwise interactions, naming pictured objects to coordinate with their partner. Successful coordination was rewarded, while failure was penalized. The goal for the participants was pairwise coordination, not global consensus. Participants had no information about the population size or connections. When a convention was established among all participants, a small number of committed individuals were introduced who continued entering the alternative strategy until the game was complete. The size of the committed minorities ranged from 15\% to 35\%. They showed that approximately 25\% of committed individuals can change the established social convention, as illustrated in Figure \ref{f02}, which shows the final success levels for overturning the established social conventions across all their trials. Furthermore, the critical mass size of approximately 25\% was found to be robust against changes in the population size, and the existence of the tipping point was also robust against the memory length.


\subsection{Color coordination game} In the color coordination game\footnote{We cover the networked color game because, although it does not strictly adhere to the traditional game-theoretical framework, it is a close variant of coordination games.}, a group of agents, represented as nodes in a fixed network, aims to select their colors simultaneously from a given set of colors~\cite{johnson1991optimization,kearns2006experimental,judd2010behavioral,kun2013anti}. The collective goal is for each node's color to be different from its neighboring nodes. 
The minimum number of colors required to achieve a conflict-free coloring of the entire network is known as the chromatic number of the network, representing the minimum necessary color space for achieving a harmonious coloring. Kearns \etal~ carried out large-scale human behavior experiments and allowed participants to access only local information, such as their neighbors' colors and their own color~\cite{kearns2006experimental}. This experimental setting allowed the game to proceed even if an agent solved the problem from their individual viewpoint. The success of the game was measured by the number of remaining conflicts within a given time length. If groups of agents can reach a global solution within the designated time, they received a bonus. Conversely, if the game is not resolved within the designated time, no bonus is awarded. This experimental model setup serves as a foundational framework for investigating the networked coloring game. In line with this human-subject experiment, Judd \etal~ conducted extensive behavioral experiments and revealed a strong correlation between individual behavioral characteristics. In particular, individuals displaying resistance to changing their initial preferences or decisions (i.e., stubborn individuals) tended to earn higher amounts~\cite{judd2010behavioral}. Additionally, Shirado \etal~ showed that including strategically placed bots in the network, programmed with a noisy strategy with which individuals choose colors that minimize conflicts with neighbors and maintain their current color otherwise, improved collective performance of human groups~\cite{shirado2017locally}. They also found that the impact of bots with small noise was comparable to assigning nodes with fixed colors, directly corresponding to committed individuals, to achieve compatibility with a global solution. Other related numerical or theoretical solutions to this problem include reinforcement learning algorithms ($q$-bots) or myopic artificial agents~\cite{qi2019social, jones2021random}. Although these studies employed more sophisticated algorithms than simply designing agents with  colors, the notion of committed individuals persists, as these preprogrammed bots maintained consistent strategies without adaptation.
\section{Social dilemma game}
Contrary to the coordination problem, the cooperation problem necessitates individuals making sacrifices for the collective good. In the following, we summarize the recent progresses on the impact of committed individuals in cooperation. 

\subsection{One-shot social dilemma games}
The basic form of the social dilemma game involves two players and two strategies: cooperation (C) and defection (D). In this game, mutual cooperation results in a reward payoff $R$, while mutual defection leads to a punishment payoff $P$. When a cooperator interacts with a defector, the former receives a sucker's payoff $S$, and the latter obtains a temptation payoff $T$. These parameters create four types of two-player social dilemma games, classified as follows \cite{wang2015universal}: prisoner's dilemma game (PD: $T>R>P>S$), snowdrift game (SD: $T>R>S>P$), stag hunt game (SH: $R>T>P>S$), and harmony game (H: $R>T, S>P$).

In one-shot social dilemma games, where reciprocity mechanisms such as reputation effects~\cite{XIA20238,wang2023reputation}, repeated interactions, and other factors like prior commitment and intention are excluded, the committed player can be either a committed cooperator or a committed defector due to the limited strategy space in this framework. The concept of committed individuals in social dilemma games was initially introduced by Mauro Mobilia~\cite{PhysRevE.86.011134}. He proposed cooperation facilitators, which enhance the fitness of cooperators but have no effect on defectors' fitness. Additionally, cooperation facilitators do not accumulate payoffs, and their strategies cannot spread. Mobilia's research focused on cooperators in finite populations and provided analytical conditions for the fixation of cooperation~\cite{PhysRevE.86.011134,MOBILIA2013113}. Later, the impact of cooperation facilitators was extended to networked populations~\cite{PhysRevE.89.042802}. Around the same time, Naoki Masuda introduced zealots in one-shot social dilemma games. Zealots enhance the payoff of both cooperators and defectors and can attract others to adopt their strategy, which distinguishes them from cooperation facilitators~\cite{masuda2012evolution}. Masuda's study explored imperfect zealous cooperators who occasionally choose defection, finding that cooperation prevails under certain conditions~\cite{masuda2012evolution}. Later, Masuda and his collaborators extended their results on infinite and well-mixed populations to finite well-mixed populations \cite{nakajima2015evolutionary}. In finite populations, a first quantity to be looked at is the fixation probability, which is the probability that a given strategy eventually dominates the population as a result of stochastic evolutionary games. The presence of zealots makes the fixation trivially occur since the zealots' strategy, assuming that it is the same for all the zealots, always fixates. Therefore, they focused on the fixation time in the presence of zealots and showed that, unlike the case without zealots, there is a threshold selection intensity below which the fixation is fast for an arbitrary payoff matrix. Matsuzawa \etal~ investigated the influence of zealots on cooperation within structured populations and discovered that committed individuals do not consistently promote cooperation when a spatial structure is introduced~\cite{matsuzawa2016spatial}. Cardillo and Masuda investigated how and why zealots can trigger the large-scale critical mass of cooperators in both well-mixed and structured populations~\cite{cardillo2020critical}. They found that only the stag-hunt game displays a clear critical mass effect, while the critical mass of cooperation in the other two dilemmas (snowdrift game and prisoner's dilemma game) relies on the selection pressure, the updating rule, or considers heterogeneous networks.

\begin{figure}
\centering
\includegraphics[scale=5]{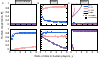}
\caption{Fraction of cooperation at stable state in well-mixed populations under the pairwise comparison rule. Each column, from left to right, represents the results for the Prisoner's Dilemma ($PD$, $S=-0.2$, $T=1.3$), Snowdrift ($SD$, $S=0.5$, $T=1.2$), and Stag Hunt ($SH$, $S=-0.2$, $T=0.6$) games. The lines depict theoretical outcomes, while the squares represent agent-based simulation results. Solid lines correspond to stable equilibria, while dashed lines represent unstable equilibria. The top and bottom panels correspond to scenarios with imitation strength parameters $\phi=0.1$ and $\phi=0.9$, respectively. We vary the imitation strength parameter $K$, where $K=100$ and $K=10$ represent strong imitation strength scenarios, and $K=1$ represents a weak imitation strength scenario. Adapted from reference ~\cite{guo2023facilitating}.}
\label{f03}
\end{figure}

Guo \etal~ made two key extensions based on the aforementioned work by Cardillo \etal ~\cite{cardillo2020critical} in the context of human-machine hybrid populations~\cite{guo2023facilitating}. Firstly, they relaxed the assumption that a committed player always chooses a specific strategy. They instead employed a probability $\phi \in [0,1]$ of selecting strategies to control the behavior of committed individual, where $\phi=1$ represents a noiseless committed individual. Secondly, they introduced a substantial number of committed players as autonomous agents. They found that cooperative committed individuals have a limited impact in the prisoner’s dilemma game (see Figure \ref{f03}A), but they facilitate cooperation in stag hunt games, and defective committed individuals, which initially seem detrimental to cooperation, can promote the complete dominance of cooperation in snowdrift games. However, as the proportion of committed individuals substantially increases, both cooperative and defective committed individuals can lead to transitioning from a coexistence of cooperators and defectors to a unanimity of defectors (see Figure \ref{f03}B and C). Moreover, in scenarios with weak imitation strength, cooperative committed individuals are able to maintain or even promote cooperation in all these games (see bottom panels of Figure \ref{f03}). Recently, Wang \etal~ studied the public goods game with committed individuals~\cite{wang2023zealous}. They found that, in spatial populations, a moderate number of committed cooperators can diminish overall cooperation, and cooperation is promoted only when committed cooperators are numerous. However, in unstructured populations, committed cooperation always promotes cooperation, indicating that the hindrance effect is primarily a spatial phenomenon. Majhi investigated how a small fraction of the committed cooperators and defectors governs the evolutionary game dynamics in which interactions among individuals with the same strategy are more probable than random interactions \cite{majhi2023cooperation}. Zhang introduced committed cooperators and defectors into the spatial prisoner's dilemma game simultaneously and found that a minority of committed players could inhibit cooperation, but the presence of noise could cancel this inhibition effect \cite{zhang2022effects}. Furthermore, the concept of committed individuals has been applied in both the ultimatum game to address fairness concerns~\cite{santos2019evolution,zheng2022probabilistic} and in AI-related research within the field of evolutionary game theory~\cite{cimpeanu2022artificial,booker2023discriminatory}.

In contrast to the aforementioned works that primarily focused on social dilemma games with two basic strategies, there are also several studies that have extended the research on committed individuals to social dilemma games with multiple strategies, addressing various problems. For instance, in the conundrum of punishment, where the effectiveness of prosocial punishment on cooperation is challenged by second-order free riders and antisocial punishment~\cite{hauser2014punishment}, Shen \etal~ attempted to address this issue in the context of human-machine hybrid populations~\cite{shen2022simple}. The authots used a two-stage prisoner's dilemma game, where players needed to choose between cooperation and defection in the first stage and decide whether to punish their opponent in the second stage. The introduction of simple bots, designed as committed prosocial punishers, allowed prosocial punishment among human players to dominate in both well-mixed and networked populations. Furthermore, when human players exhibited a learning bias towards the ``copy the majority'' rule or the bots were placed at higher-degree nodes in scale-free networks, full dominance of prosocial punishment remained possible even at high dilemma strength scenario. These results provided a novel explanation for the evolution of prosocial punishment. In line with the framework of human-machine game, Sharma \etal~ explored how simple bots, designed based on the concept of committed individuals, could facilitate the evolution of cooperation in optional prisoner's dilemma games\cite{sharma2023small}. The model involved three strategies: cooperation, defection, and loner, and two types of players: human players and bot players.  There were four pre-designed bot strategies: always choosing cooperate, always choosing defect, always choosing to be a loner, and choosing cooperation, defection, and loner randomly.  Results showed that cooperative bots promoted cooperation in well-mixed populations and regular lattices under weak imitation scenarios. Moreover, introducing loner bots enhanced cooperation in regular lattices under strong imitation scenarios, but an excess of loner bots hindered cooperation.

\subsection{Social dilemma game with repeated interactions} 
Shirado and Christakis expanded the research scope on committed individuals from one-shot social dilemma games to repeated interactions, enabling players to engage with opponents multiple times~\cite{shirado2020network}. This expanded framework introduced diverse strategies like tit-for-tat, which begins with cooperation and mirrors the opponents' actions in the subsequent rounds\cite{nowak1992tit}, and extortion, establishing a linear relationship between an individual's payoff and their opponent's payoff~\cite{press2012iterated,zhu2022nash}. In contrast, one-shot social dilemma games are limited to ALL C (cooperate) and ALL D (defect) strategies. This strategy diversity enables a more flexible design of committed strategies. Shirado \etal~ investigated how simple bots influence cooperation in repeated prisoner's dilemma games~\cite{shirado2020network}. The bots consistently cooperated with participants and allowed them to rewire connections. They showed that bot intervention significantly altered interaction dynamics, ultimately fostering sustained cooperation among participants. This finding highlights the potential of simple AI bots to enhance cooperation within a group, effectively addressing the challenges posed by the public goods dilemma. He \etal~ studied whether and how committed individuals can trigger the critical mass of cooperation in repeated games,  ~\cite{he2023impact}, and found that in scenarios with weak imitation, a committed minority of unconditional cooperators could effectively induce full cooperation. Conversely, in scenarios with strong imitation, a committed minority of conditional cooperators employing extortion strategies could encourage widespread cooperation. Remarkably, these findings remained consistent across various network topologies and imitation rules, suggesting the existence of a universal principle governing critical mass effects in social dilemma games.

\subsection{Vaccination game} 
In contrast to the pure social dilemma game, the vaccination game combines the social dilemma with disease-spreading processes~\cite{fu2011imitation}. The dilemma arises from the conflict between personal and collective interests: individuals must decide whether to avoid vaccination or take the vaccine for the maximum collective benefit. For a rational individual, the best option is to avoid vaccination, regardless of what others do, because free riders who pay nothing for vaccination can still escape infection if herd immunity is established. Liu \etal~ investigated the impact of committed vaccinated individuals who always hold a vaccination strategy on the vaccination rate and disease spreading~\cite{liu2012impact}. They found that a small fraction of these committed individuals, acting as `steadfast role models' in the population, can inhibit the formation of clusters of susceptible individuals and help promote vaccination coverage, thus inhibiting disease spreading. 
Fukuda and Tanimoto relaxed the assumption of solely committed vaccinated individuals, instead, they investigated the implications of having both committed vaccinated and unvaccinated individuals coexist within a social network under a voluntary vaccination policy~\cite{fukuda2016effects}. They revealed that the presence of committed unvaccinated individuals does not undermine the positive impact of committed individuals on collective vaccination behavior. In fact, a small fraction of committed vaccinators can still effectively promote vaccination behavior and hinder epidemic spread, even when committed unvaccinated individuals are present. 

\section{Conclusions and prospects}
In conclusion, this survey underscores the profound impact of committed individuals on coordination and social dilemma games. Committed individuals play a significant role in accelerating the emergence or overturning of social conventions, expediting solutions for color coordination issues, addressing cooperation and fairness challenges, resolving vaccination dilemmas, among others. Furthermore, these influences are shaped by players' internal properties, such as memory length, imitation strength, or strategy updating rules, as well as external factors, including network topologies, population size, or the game model. While significant progress has been made, there remain promising avenues for future exploration.

\subsection{Empirical evidence on the impact of committed individuals}
While theoretical studies have shown how and why committed individuals can influence social dynamics in various aspects, the empirical validation of their potential impact within the framework of evolutionary game theory poses significant challenges. To address this critical gap, one promising avenue involves the identification of committed individuals in both online and offline human behavior data and the subsequent investigation of their potential influence on social conventions and cooperation. Furthermore, theoretical studies have demonstrated that committed individuals can trigger a large-scale critical mass effect of cooperation~\cite{cardillo2020critical,he2023impact}. However, whether these theoretical findings can be applied to human society remains an open question. Bridging this gap necessitates the deployment of human behavior experiments as powerful tools. Not only do these experiments offer invaluable insights for reevaluating theoretical findings, but they also contribute significantly to the refinement of theoretical assumptions. As we move forward, these promising directions are poised to deepen our understanding of the mechanisms underpinning the influence of committed individuals on human behavior in real-world scenarios.

\subsection{Large language model-powered agents as proxy of committed individuals}
While theoretical studies offer valuable insights into social dynamics, one critique of previous theoretical models is that they might oversimplify the behavior of committed individuals, potentially missing the intricacies of real-world human decision-making and interactions. Recent advances in artificial intelligence, especially large language models, offer a promising avenue~\cite{grossmann2023ai}. These models, trained on massive datasets of human language~\cite{bubeck2023sparks}, are capable of simulating 
believable human behaviors \cite{park2023generative} and generating texts that are largely indistinguishable from those generated by humans \cite{brown2020language}.
We propose that these models can be used as surrogates for committed individuals in complex scenarios. For instance, in temporal networks with changing opponents~\cite{xu2023higher}, or in situations involving communication, handcrafted rules often inadequately capture committed individuals' behavior in intricate circumstances or their discourse when facing communication challenges. By employing large language models with well-crafted prompts to define committed individuals' behavior and discourse, we gain access to highly intelligent responses surpassing handcrafted rule limitations. This approach enables a more nuanced and comprehensive exploration of committed individuals' roles in diverse social dynamics.


\subsection{Addressing potential negative impacts of committed individuals}Most existing studies have focused on the positive impact of committed individuals in solving various realistic problems. Nevertheless, it is equally important to recognize the potential drawbacks associated with such individuals. For instance, during the COVID-19 pandemic, we witnessed the adverse effects of the anti-vaccination movement~\cite{fasce2023taxonomy}. This prompts a crucial inquiry: how do these detrimental influences impact society, and when do societies veer towards harmful extremes? By referring to our real world problems concerning 'virtual society' such as Twitter, SNS, and other media, it is quite important how the evolutionary game theory is able to model the committed individuals who are sometimes steadfast and stubborn intentionally leading the shared opinion to a certain direction. If such committed individuals in a virtual community are artificially generated as so-called 'bot' players, someone equipped with strong AI systems can favourably manipulate the community, which may threaten the free speech of democracy. While eliminating social media bots entirely remains a formidable challenge, mitigating the negative impact of committed individuals requires robust algorithms to detect bot accounts and strike a balance between their positive and negative aspects. A deeper understanding of the overall impact of committed individuals in social dynamics is essential to address these ethical risks effectively. 

\acknowledgments
We extend our heartfelt thanks to Prof. Naoki Masuda and Prof. Serge Galam for their thorough review of our manuscript and for offering invaluable comments that have greatly enhanced the quality of this survey. This research was supported by the National Natural Science Foundation of China (grant no. 11931015, 12271471), National Philosophy and Social Science Foundation of China (Grants No. 22\&ZD158, 22VRC049).  We also acknowledge support from (i) a JSPS Postdoctoral Fellowship Program for Foreign Researchers (grant no. P21374), and an accompanying Grant-in-Aid for Scientific Research from JSPS KAKENHI (grant no. JP 22F31374) to C.\,S., (ii) the National Science Fund for Distinguished Young Scholars (grants no. 62025602), National Natural Science Foundation of China (grants no. U22B2036, 11931015), Key Technology Research and Development Program of Science and Technology-Scientific and Technological Innovation Team of Shaanxi Province (Grant No. 2020TD-013) and the XPLORER PRIZE. to Z.\,W, and (iii) the grant-in-Aid for Scientific Research from JSPS, Japan, KAKENHI (grant No. JP 20H02314) awarded to J.\,T.

\emph{Data availability statement}: No new data were created or analysed in this study.
\bibliographystyle{eplbib}
\bibliography{ref}

\end{document}